\documentclass{article}
\usepackage[preprint]{spconfa4}
\usepackage{amsmath,amssymb,graphicx, bm}
\usepackage{hyperref}
\usepackage{booktabs}

\usepackage[skip=1pt]{caption} 

\copyrightnotice{978-1-6654-6867-1/22/\$31.00~\copyright2022 IEEE}
                   
\title{Learnable Acoustic Frontends in Bird Activity Detection}
\name{Mark Anderson, Naomi Harte}
\address{SIGMEDIA Lab \\
    School of Engineering \\
    Trinity College Dublin, Ireland\\
    \{andersm3, nharte\}@tcd.ie
}

\usepackage{tikz}
\usepackage{textcomp}
\usepackage{hyperref}
\usepackage{lipsum}
\renewcommand{\fboxsep}{2pt}
\newcommand\copyrighttext{%
  \footnotesize \textcopyright 2022 IEEE. Personal use of this material is permitted.
  Permission from IEEE must be obtained for all other uses, in any current or future
  media, including reprinting/republishing this material for advertising or promotional
  purposes, creating new collective works, for resale or redistribution to servers or
  lists, or reuse of any copyrighted component of this work in other works.
}
\newcommand\copyrightother{%
\begin{tikzpicture}[remember picture,overlay,every node/.style={inner sep=5,outer sep=20}]
\node[anchor=south, yshift=10pt, clip=false] at (current page.south) {\fbox{\parbox{\dimexpr\textwidth-\fboxsep-\fboxrule\relax}{\copyrighttext}}};
\end{tikzpicture}%
}
\begin{document}
%
\maketitle
\copyrightother{}
\begin{abstract}
Autonomous recording units and passive acoustic monitoring present minimally intrusive methods of collecting bioacoustics data. Combining this data with species agnostic bird activity detection systems enables the monitoring of activity levels of bird populations. Unfortunately, variability in ambient noise levels and subject distance contribute to difficulties in accurately detecting bird activity in recordings. The choice of acoustic frontend directly affects the impact these issues have on system performance. In this paper, we benchmark traditional fixed-parameter acoustic frontends against the new generation of learnable frontends on a wide-ranging bird audio detection task using data from the DCASE2018 BAD Challenge. We observe that Per-Channel Energy Normalization is the best overall performer, achieving an accuracy of $89.9\%$, and that in general learnable frontends significantly outperform traditional methods. We also identify challenges in learning filterbanks for bird audio.
\end{abstract}
\begin{keywords}
Bird Activity Detection, Bioacoustics, Learnable Frontend, PCEN, LEAF, STRF, TD Filterbanks
\end{keywords}
\section{Introduction}

\label{sec:intro}
Monitoring of bird populations is of increasing importance as biodiversity continues to decline and the impacts of climate change become more evident \cite{Tilman201773}. Many population studies are utilising passive acoustic monitoring and autonomous recording units to record bioacoustic audio \cite{RempelARU, biosciARU}, which present non-intrusive methods for carrying out research on bird populations \cite{kitzes_schricker_2019}. The results of these studies can serve as indicators of the effects of climate change \cite{FIEDLER2009181}. 

Audio is highly suitable for monitoring bird populations as many species are identifiable by their vocalisations and visual analysis can be difficult. In the age of Deep Learning, spectrogram representations of the audio are the most common input features to a system \cite{Stowell2022}. This form of representation is easily interpretable by humans and also allows Convolutional Neural Networks to extract spectro-temporal patterns. 

The primary parameters governing the spectrogram representation of audio are the window size (determining the time/frequency trade-off), frequency axis scaling and amplitude compression. There is evidence that increased time resolution favours bird audio \cite{knight_spectrograms} but there is little consensus on the best representation of the frequency axis, whether it is linear, logarithmic or mel scaling \cite{Stowell2022}. There is further debate on whether mel scaling is suitable for bioacoustics, as it is derived from human auditory perception \cite{melhuman}. Bird audio is typically in the range of 800-8000Hz, and mel scaling focuses much of the energy in the lower bands.  Log compression is loudness dependant and can increase the appearance of noise in the spectrogram. The robustness of the chosen representation to environmental noise and loudness variation is paramount. In spite of this, standard log-mel spectrograms used in human speech applications are ubiquitous in the analysis of bird vocalisations \cite{Stowell2022}. 

\vspace{-.15em}
New trainable frontends have been developed for a range of audio applications which can learn the filterbanks \cite{TD, LEAF}, perform loudness normalisation and noise reduction \cite{pcen_original, LEAF}, or exploit patterns in temporal and frequency modulation \cite{strf}. Per-Channel Energy Normalisation has been employed recently in bioacoustics \cite{Lostanlen2019, pcenSEDbio} and the remaining frontends have been tested on various bird audio tasks, including bird activity detection. However, no dedicated comparative assessment of the suitability of these methods in bird activity detection has been done. As a vital first step to any subsequent population analysis, this is crucial for judging their suitability for bird bioacoustics work.

\vspace{-.15em}
This paper compares traditional and learnable frontends using the same datasets and model architecture. We evaluate performance on a species-agnostic bird activity detection task. The datasets contain a large variety of bird species and call types, as well as many noise sources (e.g. wind, traffic, human activity and speech). We provide the most comprehensive, up-to-date investigation on the suitability of learnable frontends in bird audio detection to date. Section \ref{sec:frontends} details the frontends. In Section \ref{sec:method} we outline our methodology for evaluation. In Section \ref{sec:results}, we present and discuss the experimental results and Section \ref{sec:conc} concludes the paper.

\section{Frontends for Bird Audio Detection}
\label{sec:frontends}
Frontends in this paper can be broken into two groups: non-learnable, and learnable. Non-learnable representations include linear spectrograms (spect), mel spectrograms (mel), and log compressed mel spectrograms (logmel). These frontends have no parameters that are optimised during network training; their parameters are fixed.  Our choice in non-learnable frontend parameters was influenced by \cite{knight_spectrograms} and \cite{graciarena}. We utilise a window length of $10ms$ in all our FFTs with overlap of $75\%$, and 41 mel filters with center frequencies ranging from 500Hz to 16KHz.

The learnable frontends include Spectro-Temporal Filters (STRF) \cite{strf}, Time-Domain Filterbanks (TD) \cite{TD}, Per-Channel Energy Normalisation (PCEN) \cite{pcen_original} and LEarnable Audio Frontend (LEAF) \cite{LEAF}. These frontends represent a new generation of frontends implemented as layers in a neural network, which can be optimised alongside the model. We provide a brief explanation of each learnable frontend below.

\subsection{STRF}
STRF was proposed by Riad et al. \cite{strf} in 2021 as a general purpose acoustic frontend. It comprises a set of learnable two-dimensional Gabor filters which detect spectro-temporal modulations. The primary focus on temporal and spectral modulations has been utilised previously in low computation bird activity detection \cite{amfmAnderson}. The authors limit the frontend to the space of Gabor functions $g(t, f)$, defined by Equation \ref{eqn:strf_filter}, which comprises of a sinusoidal wave $s(t, f)$ (Equations \ref{eqn:strf_sine}, \ref{eqn:strf_R}) windowed by a Gaussian function $w(t, f)$ (Equation \ref{eqn:strf_window}). 

\vspace{-1.5em}
\begin{align}
    g_{k}(t,f) &= s_{k}(t,f) \cdot w_{k}(t,f) \label{eqn:strf_filter} \\
    w_{k}(t,f) &= \frac{1}{2\pi\sigma_{t_k}\sigma_{f_k}} e^{-(\frac{1}{2}(t^2/\sigma^2_{t_k}+f^2/\sigma^2_{f_k}))} \label{eqn:strf_window} \\
    s_{k}(t,f) &= e^{j(2\pi(F_k R_{\gamma_k}))} \label{eqn:strf_sine} \\
\text{where,\hspace{2em}} 
    R_{\gamma_k} &= t\cos(\gamma_k) + f\sin(\gamma_k) \label{eqn:strf_R}
\end{align}

STRF takes in a mel-spectrogram ($E(t,f)$) and computes 64 Gabor filters $\boldsymbol{Z}$ (Equation \ref{eqn:strf_overall}) with four learnable parameters per filter controlling the sinusoidal wave ($F_k, \gamma_k$) and Gaussian window ($\sigma_{t_k}, \sigma_{f_k}$) of each filter. 

\vspace{-1em}
\begin{align}
    \textbf{Z}(t,f,k) &= \sum_{u,v}^{}E(u,v)g_k(t-u,f-v) \label{eqn:strf_overall}
\end{align}
\vspace{-1em}

STRF has been tested on a zebra finch call classification task (single species, distinguishing between different vocalisations), but has not been tested on our task of species-agnostic bird activity detection.

\subsection{TD}
TD was proposed by Zeighdour et al. \cite{TD} in 2018 as an alternative to mel filterbanks, proposing a set of learnable filterbanks tuned to the specific application. There are also options for the learning of pre-emphasis and averaging.

\vspace{-1.1em}
\begin{align}
    \text{TD}(t,f) &= \lvert x \ast \phi_n \rvert^2 \ast \lvert\Phi\rvert^2(t) \label{eqn:td}
\end{align}
\vspace{-1.1em}

Equation \ref{eqn:td} defines the time-frequency representation of TD, where $x$ is the input waveform, $\phi_n$ is the impulse response of the $n$-th filter and $\Phi(t)$ represents the Hanning window. The wavelets ($\phi_n$) are initialised to Gabor wavelets approximating a mel filterbank. Although initialised to an approximate mel scale using Gabor wavelets, the parameters are not constrained to this. 

TD has been tested on a bird audio detection task \cite{LEAF} on the same data utilised in this paper, achieving an accuracy of $80.9\%$. Our experiments using TD outperform the results presented in \cite{LEAF}.

\subsection{PCEN and LEAF}
PCEN (Wang et al.) \cite{pcen_original} and LEAF (Zeghidour et al.) \cite{LEAF} were both proposed by researchers at Google in 2017 and 2021, respectively. PCEN was originally proposed to improve keyword spotting, whereas LEAF is a general-purpose acoustic frontend. PCEN usually takes mel-spectrograms as input ($E(t,f)$) but there is no restiction on what filterbanks could be used. It addresses the issues of normalisation and noise by proposing learnable Automatic Gain Control (AGC) and Dynamic Range Compression (DRC) parameters. The AGC is applied prior to DRC and yields Equation \ref{eqn:pcen}. Both the AGC and DRC are learned per frequency channel.

\vspace{-1em}
\begin{align}
    \text{PCEN}(t,f) = \left(\frac{E(t,f)}{(M(t,f) + \epsilon)^\alpha} + \delta\right)^r -\delta^r \label{eqn:pcen}
\end{align}
\vspace{-1em}

AGC is achieved using a learnable smoother (Equation \ref{eqn:pcen_smoother}) which emphasises changes relative to the recent spectral history along the temporal axis. The smoothing parameter $s$ can be learned for each frequency band, as a global parameter or set according to some time constant $T$ \cite{PCENwhy}. Therefore PCEN is based on the set of parameters $(\boldsymbol{s}, \boldsymbol{\alpha}, \boldsymbol{\delta}, \boldsymbol{r}, \epsilon)$

\vspace{-1.2em}
\begin{align}
    M(t,f) = (1 - s)M(t-1,f) + sE(t,f) \label{eqn:pcen_smoother}
\end{align}
\vspace{-1.2em}

The overall effect is one of Gaussianising the distribution of spectral magnitudes and decorrelating frequency bands \cite{PCENwhy}. The output of PCEN is a spectrogram-like output and has already been employed in bioacoustic Sound Event Detection to great effect \cite{Lostanlen2019, pcenSEDbio}. 

LEAF extends upon PCEN through a learnable bank of one-dimensional Gabor filters and low-pass smoothing (similar in formulation to Equation \ref{eqn:td}) prior to applying PCEN. The Gabor filters are initialised to the mel scale at the start of training and the system attempts to learn the frequency bands of interest. Unlike TD, these filters are limited to the space of Gabor functions (Equation \ref{eqn:leaf_gabor}). This results in fewer parameters than the unconstrained filters used in TD, and the filterbanks are more easily interpretable.

\vspace{-1.5em}
\begin{align}
    \phi_n(t) &= e^{2\pi j\eta_n t}\frac{1}{\sqrt{2\pi}\sigma_{n_{bw}}}e^{-\frac{t^2}{2\sigma^2_{n_{bw}}}} \label{eqn:leaf_gabor} \\
    \Phi_n(t) &= \frac{1}{\sqrt{2\pi}\sigma_{n_{lp}}}e^{-\frac{t^2}{2\sigma^2_{n_{lp}}}} \label{eqn:leaf_lowpass}
\end{align}
\vspace{-1em}

LEAF uses learnable low-pass filters for each frequency band in order to downsample in the time-domain. In contrast to TD, which uses a Hanning window, LEAF utilises a low pass filter with a Gaussian impulse response (Equation \ref{eqn:leaf_lowpass}). This produces a spectrogram like-output similar to PCEN. In addition to the parameters for PCEN, LEAF adds the following parameters $(\eta_n, \sigma_{n_{bw}}, \sigma_{n_{lp}})$. LEAF has achieved an accuracy of $81.4\%$, also on the bird audio detection tasks in \cite{LEAF}.

\section{Testing Methodology}
%
%
\label{sec:method}
\begin{table}[t]
\centering
\small
\label{tab:dataset}
\begin{tabular}{l|lll}
\hline
\textbf{Dataset}  & \textbf{Positive} & \textbf{Negative} & \textbf{Total} \\ \hline
BirdVox-DCASE-20k & 10017             & 9983              & 20000          \\
freefield1010     & 5755              & 1935              & 7690           \\
warblrb10k        & 6045              & 1955              & 8000           \\ \hline
\textbf{Totals}   & 21817             & 13873             & 35690       
\end{tabular}
\caption{Details of datasets included from the DCASE2018 Bird Audio Detection Challenge. Positive means count of recordings with a bird present.}
\end{table}
\subsection{Task \& Datasets}
\label{sec:datasets}
To evaluate each frontend we train independent supervised models on a bird audio detection task. This task involves the classification of whether a 10s clip of audio contains bird vocalisations or not. It is a species agnostic task, and the system is expected to generalise to species and vocalisations it may not have encountered during training.

We utilise the datasets released as part of the DCASE2018 Bird Audio Detection Challenge \cite{DCASE2018}. The challenge provided annotated datasets from three sources (detailed in Table 1), in order to provide a better evaluation of generality. One dataset (\textit{BirdVox-DCASE-20k} \cite{birdvox-dcase}) contains passively recorded data from remote monitoring projects. The remaining datasets (\textit{freefield1010} \cite{ff1010}, \textit{warblrb10k}\footnote{Accessible via: \href{https://archive.org/details/warblrb10_public}{https://archive.org/details/warblrb10k\_public}}), are active, crowdsourced recordings contributed to the freesound project and from users of the Warblr bird recognition app.

All three datasets combined comprise 35690 recordings, each of 10s length, sampled at 44.1KHz and normalised to $-2\text{dBFS}$. Clip-level annotations are provided denoting the presence (postive label) or absence (negative label) of birds. BirdVox is considered a more challenging dataset due to the abundance of environmental noise and distance from subject. Distance to subject is shorter in Freefield1010 and Warblrb10k however, both contain large amounts of human generated noise.  We use a 70:15:15 split for training, validation and test datasets, ensuring the relative proportions and class balance of each dataset are maintained. 

\subsection{Model}
The network architecture for all our experiments is EfficientNet-B0 \cite{efficientnet} with one head node providing classification output. The most popular model architectures for bioacoustics work are ResNet or VGG-based \cite{Stowell2022}. However, EfficientNet based models offer a good compromise between computational resources and accuracy, and can be deployed on edge-based systems. We do not use pre-trained weights to initialise the network. We train using Binary Cross Entropy loss and the ADAM optimiser with an initial learning rate of $10^{-3}$ with a $10$x reduction when the validation loss reaches a plateau.
\subsection{Statistical Testing}
\label{sec:analysis}
We analyse our results by first reporting the accuracy of the model using a particular frontend, evaluated across the entire test set. We also report accuracy by dataset. We take random, independently distributed subsets with replacement of the test set, and evaluate the accuracy of each subset per model. The Shapiro-Wilk test of normality was used and indicated that the data are normally distributed and the results of ANOVA testing indicate at least one of the systems has statistically different results. Verifying this allows us to run a post-hoc pairwise comparison. 

As our data is normally distributed, we use the Tukey HSD test to find which pairs of results differ from each other. This determines whether results are significantly different. If there is no significant difference between the two frontends on the whole test set, then further analysis will be carried out by comparing their accuracy per dataset.
    
\section{Results \& Discussion}
\label{sec:results}


\begin{table}[t]
\small
\centering
\label{tab:result}
\begin{tabular}{l|c}
\hline
\textbf{Frontend} & \multicolumn{1}{l}{\textbf{Test Set Accuracy (\%)}} \\ \hline
\textbf{spect}    & 78.4                                                \\
\textbf{mel}      & 71.7                                                \\
\textbf{logmel}   & 70.4                                                \\
\textbf{STRF}     & 71.3                                                \\
\textbf{TD}       & 87.6                                                \\
\textbf{PCEN}     & \textbf{89.9}                                       \\
\textbf{LEAF}     & 83.7                                               
\end{tabular}
\caption{Test accuracy, broken down by frontend.}
\end{table}

\begin{table}[t]
\centering
\small
\label{tab:tukey}
\begin{tabular}{c|ccccccc}
\hline
 & spect & mel & logmel & STRF & TD & PCEN \\\hline
spect & N/A &  &  &  &  & \\
mel & $\blacksquare$  & N/A &  &  &  & \\
logmel & $\blacksquare$ & $\square$ & N/A &  &  & \\
STRF & $\blacksquare$ & $\square$ &  & N/A &  & \\
TD & $\blacksquare$  & $\blacksquare$  & $\blacksquare$ & $\blacksquare$ & N/A & \\
PCEN & $\blacksquare$ & $\blacksquare$ & $\blacksquare$ & $\blacksquare$ & $\blacksquare$ & N/A \\
LEAF & $\blacksquare$ & $\blacksquare$ & $\blacksquare$ & $\blacksquare$ & $\blacksquare$ & $\blacksquare$\\
\end{tabular}
\caption{Significance tests on pairwise-comparisons using Tukey's HSD. Cells marked with $\blacksquare$ indicate statistically significant results ($p < 0.05$) on the entire test set. Cells marked with $\square$ indicate statistically significant results on at least one dataset.}
\end{table}

\begin{figure}[t]
\centering
\includegraphics[trim=10 11 10 12,clip,width=0.45\textwidth]{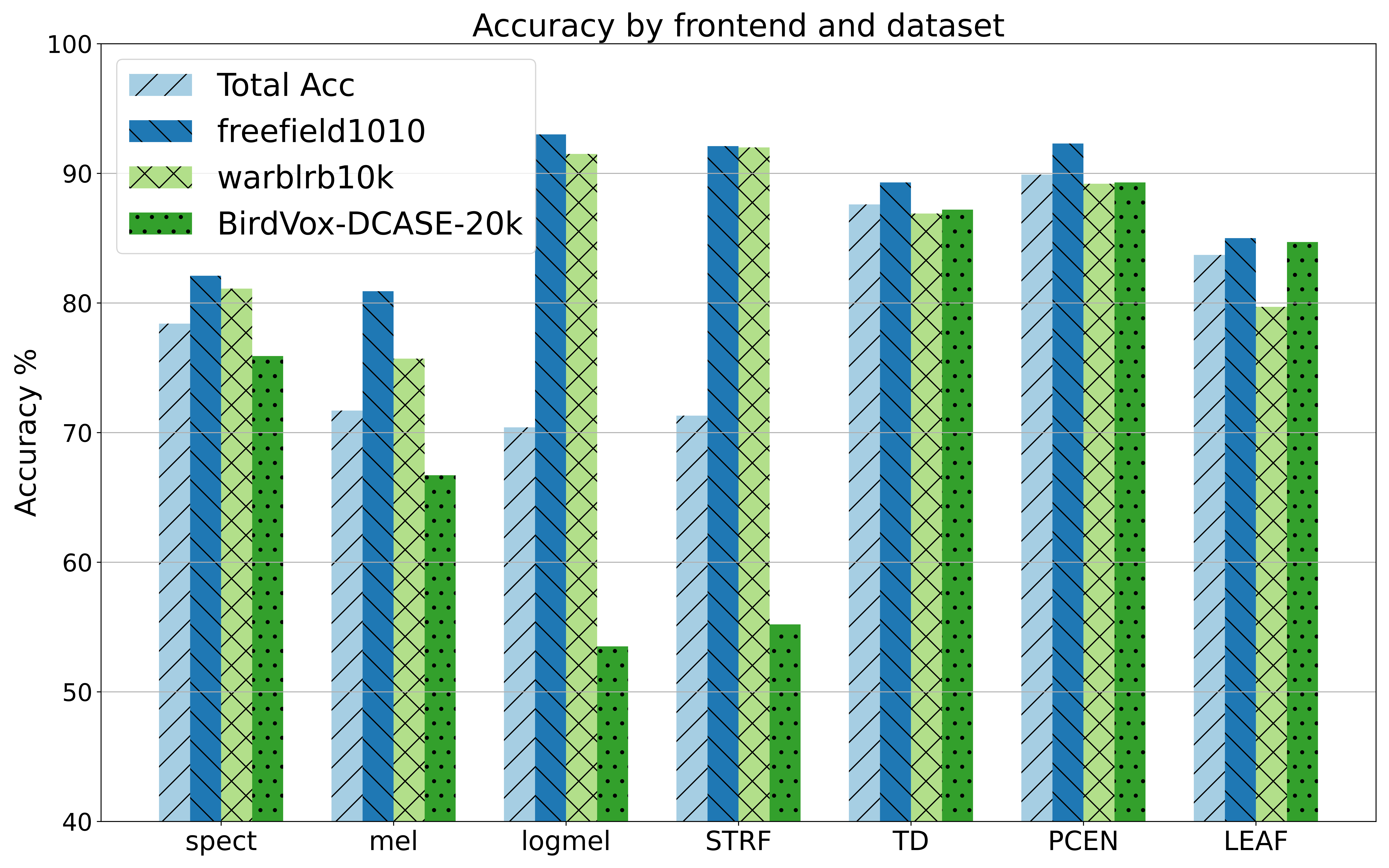}
\caption{\label{fig:results}Total Accuracy on the test set, by frontend.  Accuracy is also broken down by dataset}
\end{figure}

The overall results are presented in Table 2 and Figure \ref{fig:results}. Table 2 reports the performance of each frontend on the test data, allowing for an overall system ranking. From this we can see that PCEN is the best overall performer. Figure \ref{fig:results} details performance per dataset, providing insight into how each frontend performs under different conditions. Table 3 presents the results of significance testing. As we can see most of our results are statistically significant. Only the STRF/logmel results do not differ significantly on any dataset, implying there is no discernible performance difference.

Of the non-learnable frontends, uncompressed linearly scaled spectrograms (spect) perform best ($78.4\%$). Uncompressed mel filterbank energies (mel) are the next best ($71.7\%$) with similar performance on freefield1010, but reduced performance on warblrb10k and BirdVox-DCASE-20k. Log compressed mel spectrograms (logmel) is the worst traditional frontend ($70.4\%$), largely due to its poor performance on the BirdVox data. Logmel offers excellent performance on the cleaner datasets,  better than some learnable frontends. The log compression however, causes very poor performance on noisier data with distant subjects.

On average, the learnable frontends perform better than the non-learnable frontends. STRF ($71.3\%$) offers similar performance to logmel ($70.4\%$) and as mentioned above, Table 3 shows there is no significant difference in their results on any dataset. The reasons for this are currently unknown, as both systems operate on different principles. The only conclusion we can make at this time is that on this task, STRF performs no better than logmel and is less interpretable. TD shows the second best performance overall ($87.6\%)$, showing that learnable filterbanks are capable of improving performance. The results across all datasets are consistent, with good performance on more difficult data. However the TD representation is less interpretable, likely due to the unconstrained filterbanks.

As previously mentioned, PCEN is the best performer overall ($89.9\%$), offering similar performance gains to logmel on the cleaner data of freefield1010 and warblrb10k, but also on the more difficult data from BirdVox. This is due to the AGC being tuned to improve the dynamic range of the signal of interest, while lessening the effect of noise on the resulting spectrogram. Note that in our experiments, we could not train a per-channel smoother to further improve normalisation and noise reduction; these models could not converge. Therefore we used a fixed value for $\boldsymbol{s}$.

LEAF ($83.7\%$) performs worse than PCEN ($89.9\%$) and TD ($87.6\%$), despite incorporating elements from both approaches. It offers the advantage of learnable filters that are more interpretable than TD, and uses a PCEN layer for compression of the resulting magnitudes, however, it performs worse on all datasets. There are two possible reasons for this: the learned filterbanks or the learned lowpass filters. Given the strong performance of TD, future work should compare the learned filterbanks of TD and LEAF, to understand why LEAF did not perform best on this task. 

Our results indicate that learnable compression of spectrogram magnitudes offer the greatest increases in performance. Log compression works well on clean data, but on noisier data the performance suffers. No compression at all is more consistent than log scaling, and learnable compression is both consistent and performs better. Learnable filterbanks can improve performance, but less so than learnable compression, and in the case of PCEN vs. LEAF, seem to reduce performance. We posit that the complexity of bird audio, especially in a species agnostic task, poses issues for learning filterbanks that are constrained to a given space (i.e. the space of Gabor functions in the case of LEAF). This constraint is imposed for interpretability, but negatively impacts performance on this challenging data. Further work is needed to improve the performance of learnable filterbanks that remain interpretable.

\section{Conclusion}
\label{sec:conc}
In this paper, we compared traditional and learnable acoustic frontends for audio classification in the context of bioacoustics, specifically bird activity detection. We observe significant improvements in model accuracy using learnable frontends. PCEN is the best overall performer, and the results from LEAF and TD suggest that the effect of compression has a greater impact on system performance than the filterbanks used. While acknowledging the gains shown by learnable filterbanks over traditional methods, we recommend the usage of PCEN over other methods until issues around learnable interpretable filterbanks for bird audio detection are resolved. 
\newpage
\let\oldthebibliography=\thebibliography
\let\endoldthebibliography=\endthebibliography
\renewenvironment{thebibliography}[1]{%
   \begin{oldthebibliography}{#1}%
     \setlength{\itemsep}{+.15ex}%
}%
{%
   \end{oldthebibliography}%
}
\bibliographystyle{IEEEbib}
\bibliography{refs}

\begin{thebibliography}{10}

\bibitem{Tilman201773}
D.~Tilman, M.~Clark, D.R. Williams, K.~Kimmel, S.~Polasky, and C.~Packer,
\newblock ``Future threats to biodiversity and pathways to their prevention,''
\newblock {\em Nature}, vol. 546, no. 7656, pp. 73--81, 2017.

\bibitem{RempelARU}
R.S. Rempel, C.M. Francis, J.N. Robinson, and M.~Campbell,
\newblock ``Comparison of audio recording system performance for detecting and
  monitoring songbirds,''
\newblock {\em Journal of Field Ornithology}, vol. 84, no. 1, pp. 86--97, 2013.

\bibitem{biosciARU}
Larissa Sayuri~Moreira Sugai, Thiago Sanna~Freire Silva, Jr~Ribeiro,
  José~Wagner, and Diego Llusia,
\newblock ``{Terrestrial Passive Acoustic Monitoring: Review and
  Perspectives},''
\newblock {\em BioScience}, vol. 69, no. 1, pp. 15--25, 11 2018.

\bibitem{kitzes_schricker_2019}
Justin Kitzes and Lauren Schricker,
\newblock ``The necessity, promise and challenge of automated biodiversity
  surveys,''
\newblock {\em Environmental Conservation}, vol. 46, no. 4, pp. 247–250,
  2019.

\bibitem{FIEDLER2009181}
Wolfgang Fiedler,
\newblock ``Chapter 9 - bird ecology as an indicator of climate and global
  change,''
\newblock in {\em Climate Change}, Trevor~M. Letcher, Ed., pp. 181--195.
  Elsevier, 2009.

\bibitem{Stowell2022}
Dan Stowell,
\newblock ``Computational bioacoustics with deep learning: a review and
  roadmap,''
\newblock {\em PeerJ}, vol. 10, pp. e13152, Mar. 2022.

\bibitem{knight_spectrograms}
Elly~C. Knight, Sergio~Poo Hernandez, Erin~M. Bayne, Vadim Bulitko, and
  Benjamin~V. Tucker,
\newblock ``Pre-processing spectrogram parameters improve the accuracy of
  bioacoustic classification using convolutional neural networks,''
\newblock {\em Bioacoustics}, vol. 29, no. 3, pp. 337--355, 2020.

\bibitem{melhuman}
S.~Davis and P.~Mermelstein,
\newblock ``Comparison of parametric representations for monosyllabic word
  recognition in continuously spoken sentences,''
\newblock {\em IEEE Transactions on Acoustics, Speech, and Signal Processing},
  vol. 28, no. 4, pp. 357--366, 1980.

\bibitem{TD}
Neil Zeghidour, Nicolas Usunier, Iasonas Kokkinos, Thomas Schaiz, Gabriel
  Synnaeve, and Emmanuel Dupoux,
\newblock ``Learning filterbanks from raw speech for phone recognition,''
\newblock in {\em 2018 IEEE International Conference on Acoustics, Speech and
  Signal Processing (ICASSP)}, 2018, pp. 5509--5513.

\bibitem{LEAF}
Neil Zeghidour, Olivier Teboul, F{\'{e}}lix de~Chaumont~Quitry, and Marco
  Tagliasacchi,
\newblock ``{LEAF:} {A} learnable frontend for audio classification,''
\newblock {\em CoRR}, vol. abs/2101.08596, 2021.

\bibitem{pcen_original}
Yuxuan Wang, Pascal Getreuer, Thad Hughes, Richard~F. Lyon, and Rif~A. Saurous,
\newblock ``Trainable frontend for robust and far-field keyword spotting,''
\newblock in {\em 2017 IEEE International Conference on Acoustics, Speech and
  Signal Processing (ICASSP)}, 2017, pp. 5670--5674.

\bibitem{strf}
Rachid Riad, Julien Karadayi, Anne-Catherine Bachoud-Lévi, and Emmanuel
  Dupoux,
\newblock ``Learning spectro-temporal representations of complex sounds with
  parameterized neural networks,''
\newblock {\em The Journal of the Acoustical Society of America}, vol. 150, no.
  1, pp. 353--366, 2021.

\bibitem{Lostanlen2019}
Vincent Lostanlen, Kaitlin Palmer, Elly Knight, Christopher Clark, Holger
  Klinck, Andrew Farnsworth, Tina Wong, Jason Cramer, and Juan Bello,
\newblock ``Long-distance detection of bioacoustic events with per-channel
  energy normalization,''
\newblock in {\em Proceedings of the Detection and Classification of Acoustic
  Scenes and Events 2019 Workshop (DCASE2019)}, New York University, NY, USA,
  October 2019, pp. 144--148.

\bibitem{pcenSEDbio}
Vincent Lostanlen, Justin Salamon, Andrew Farnsworth, Steve Kelling, and
  Juan~Pablo Bello,
\newblock ``Robust sound event detection in bioacoustic sensor networks,''
\newblock {\em PLOS ONE}, vol. 14, no. 10, pp. 1--31, 10 2019.

\bibitem{graciarena}
Martin Graciarena, Michelle Delplanche, Elizabeth Shriberg, Andreas Stolcke,
  and Luciana Ferrer,
\newblock ``Acoustic front-end optimization for bird species recognition,''
\newblock in {\em 2010 IEEE International Conference on Acoustics, Speech and
  Signal Processing}, 2010, pp. 293--296.

\bibitem{amfmAnderson}
Mark Anderson, John Kennedy, and Naomi Harte,
\newblock ``Low resource species agnostic bird activity detection,''
\newblock in {\em 2021 IEEE Workshop on Signal Processing Systems (SiPS)},
  2021, pp. 34--39.

\bibitem{PCENwhy}
Vincent Lostanlen, Justin Salamon, Mark Cartwright, Brian McFee, Andrew
  Farnsworth, Steve Kelling, and Juan~Pablo Bello,
\newblock ``Per-channel energy normalization: Why and how,''
\newblock {\em IEEE Signal Processing Letters}, vol. 26, no. 1, pp. 39--43,
  2019.

\bibitem{DCASE2018}
Dan Stowell, Michael~D. Wood, Hanna Pamuła, Yannis Stylianou, and Hervé
  Glotin,
\newblock ``Automatic acoustic detection of birds through deep learning: The
  first bird audio detection challenge,''
\newblock {\em Methods in Ecology and Evolution}, vol. 10, no. 3, pp. 368--380,
  2019.

\bibitem{birdvox-dcase}
Vincent Lostanlen, Justin Salamon, Andrew Farnsworth, Steve Kelling, and
  Juan~Pablo Bello,
\newblock ``Birdvox-full-night: a dataset and benchmark for avian flight call
  detection,''
\newblock in {\em Proc. IEEE ICASSP}, April 2018.

\bibitem{ff1010}
Dan Stowell and Mark~D. Plumbley,
\newblock ``An open dataset for research on audio field recording archives:
  freefield1010,''
\newblock {\em CoRR}, vol. abs/1309.5275, 2013.

\bibitem{efficientnet}
Mingxing Tan and Quoc~V. Le,
\newblock ``Efficientnet: Rethinking model scaling for convolutional neural
  networks,''
\newblock {\em CoRR}, vol. abs/1905.11946, 2019.

\end{thebibliography}

\end{document}